\documentclass[aps,pra,twocolumn,floatfix]{revtex4-1}

\usepackage{bm,dcolumn,amsmath,graphicx}
\usepackage{epsfig}
\usepackage{nicefrac}
\def\adndt{At. Data Nucl. Data Tables}
\def\cjp{Can. J. Phys.}

\def\epl{Europhys. Lett.}

\def\jpb{J. Phys. B}

\def\jpcs{J. Phys.: Conf. Ser.}

\def\pra{Phys. Rev. A}

\def\prl{Phys. Rev. Lett.}

\def\pscr{Phys. Scr.}

\def\sci{Science}

\newcommand{\paper}{paper}

\newcommand{\eref}[1]{(\ref{#1})}

\newcommand{\Tref}[1]{Table~\ref{#1}}
\newcommand{\Fig}[1]{Fig.~\ref{#1}}

\newcommand{\E}[1]{\ensuremath{\times 10^{#1}}}
\newcommand{\cm}{\ensuremath{\textrm{cm}^{-1}}}
\newcommand{\half}{\nicefrac{1}{2}}

\newcolumntype{b}{D{(}{\ (}{-1}}  %Align on opening bracket of errors; put space before 

\begin{document}

\title{Hole Transitions in Multiply-Charged Ions for Precision Laser Spectroscopy and Searching for $\alpha$-variation}

\author{J. C. Berengut}
\author{V. A. Dzuba}
\author{V. V. Flambaum}
\author{A. Ong}
\affiliation{School of Physics, University of New South Wales, Sydney, NSW 2052, Australia}

\date{15 March 2011}


\begin{abstract}

We consider transitions of electron holes (vacancies in otherwise filled shells of atomic systems) in multiply-charged ions that, due to level-crossing of the holes, have frequencies within the range of optical atomic clocks. Strong E1 transitions provide new options for laser-cooling and trapping, while narrow transitions can be used for high-precision spectroscopy and tests of fundamental physics. We show that hole transitions can have extremely high sensitivity to $\alpha$-variation due to large relativistic corrections, and propose candidate transitions that have much larger $\alpha$-sensitivities than any previously seen in atomic systems.

\end{abstract}

\maketitle

\section{Introduction}
Recent evidence from quasar absorption spectra suggests that there may be a spatial gradient in values of the fine-structure constant, $\alpha=e^2/\hbar c$, across cosmological distance scales~\cite{webb10arxiv}. The data come from some $\sim 300$ different absorbers using two different telescopes, Keck and the Very Large Telescope, which together provide coverage of the whole sky. All existing astronomical data measuring $\alpha$-variation are consistent with the existence of a constant spatial gradient in $\alpha$ (dipole model)~\cite{berengut11jpcs}.

The existence of a spatial dipole in $\alpha$ could be confirmed using terrestrial clocks~\cite{berengut10arxiv0}. The motion of the Sun relative to the measured dipole (that is, towards a region of the Universe with larger $\alpha$) implies a time-variation of around $\dot\alpha/\alpha \sim 10^{-18} - 10^{-19}\,\textrm{yr}^{-1}$. The annual motion of the Earth modulates the signal, giving an additional shift $\delta\alpha/\alpha = 1.4\E{-20}\cos\omega t$, where $\omega$ refers to the frequency of the annual orbit. To measure this shift, one must compare two clocks with different sensitivities to $\alpha$ over the course of several years~\cite{dzuba99prl}. The best current limit on terrestrial time-variation, $\dot\alpha/\alpha = (-1.6\pm2.3)\E{-17}\textrm{yr}^{-1}$, comes from comparison of Hg$^+$ and Al$^+$ optical clocks over the course of a year~\cite{rosenband08sci}. Here the Hg$^+$ transition is strongly dependent on $\alpha$, while the Al$^+$ clock is relatively insensitive~\cite{dzuba99prl,dzuba99pra}.

To confirm the dipole model of $\alpha$-variation in the laboratory requires finding systems where the sensitivity to $\alpha$ is enhanced. The sensitivity is indicated by $q$-coefficients, usually defined by
\begin{equation}
\label{eq:q_def}
\omega = \omega_0 + q x
%x = (\alpha/\alpha_0)^2 - 1 \approx 2\frac{\alpha - \alpha_0}{\alpha_0}
\end{equation}
where $x = (\alpha/\alpha_0)^2 - 1 \approx 2(\alpha - \alpha_0)/\alpha_0$ is the change in $\alpha$ over time from its current value $\alpha_0$, which leads to a change in the frequency $\omega$. Here the atomic unit of energy, which is cancelled out in any measured ratio of frequencies, is assumed to be constant. The dependence of $\omega$ on $\alpha$ is due to relativistic corrections.
Potential clocks that have large $q$-values include optical transitions in Yb$^+$~\cite{porsev09pra} and Th$^{3+}$~\cite{flambaum09pra}, as well as the thorium ``nuclear clock''~\cite{peik03epl}, which would make use of the 7.5 eV nuclear transition in $^{229}$Th to produce a clock with a $q$-value many orders-of-magnitude larger than the optical Hg$^+$ clock~\cite{flambaum06prl,berengut09prl}. Recently, we showed that trapped, highly-charged ions could provide optical atomic clocks with much larger $q$-values than would be possible using near-neutral ions~\cite{berengut10prl}. Clocks using suggested transitions in Sm$^{14+}$, for example, would benefit from several separate contributions to $\alpha$-sensitivity: high nuclear charge $Z$, high ionization stage, and significant differences in the configuration composition of the states involved.

In this \paper\ we demonstrate that using holes can dramatically increase sensitivity to $\alpha$-variation, while retaining all of the other sources of enhancement. We show that it is possible to find hole transitions in highly ionized atomic systems that are within the range of optical lasers, and discuss the sources of the large $q$ values. Two promising systems are identified along the ionization sequences of iridium and tungsten. The Ir$^{16+}$ and Ir$^{17+}$ ions are shown to have $q$-values much larger than any that have previously been seen in atomic systems.

\section{Theory}
\begin{figure}[tb]
\includegraphics[width=0.44\textwidth]{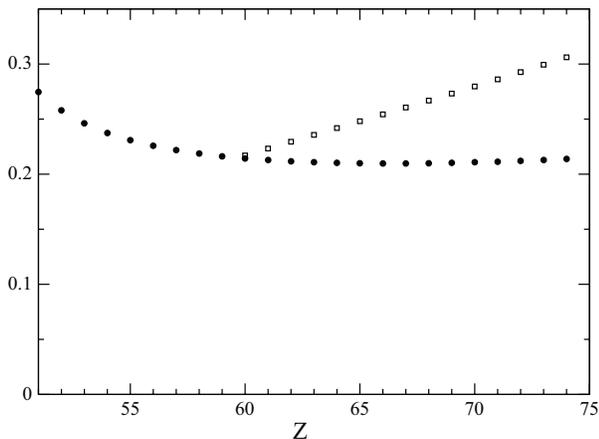}
\caption{\label{fig:q5s_norm} Ratio $q_\textrm{norm} = -q/(Z^2\alpha^2 I_n)$ for the $5s$ orbital, calculated using the Dirac-Fock theory. Filled circles: $q_\textrm{norm}$ for the $5s$ valence orbital in the Ag isoelectronic sequence. Open squares: $q_\textrm{norm}$ for a $5s$ orbital from the configuration $5s^2 4f^k$ where the ionization degree is always $Z_{ion} = 12$.}
\end{figure}

In Refs.~\cite{berengut10prl,dzuba99pra} it was shown that for a single electron above closed shells, the relativistic shift is
\begin{equation}
\label{eq:q_external_electron}
q \approx -I_n \frac{(Z\alpha)^2}{\nu(j+\half)}
\end{equation}
where $I_n = Z_a^2/2\nu^2$ is the ionisation energy of the orbital (atomic units $\hbar = e = m_e = 1$). Here $Z_a$ is the effective charge that the external electron ``sees'' and $\nu$ is the effective principal quantum number. It was shown in \cite{berengut10prl} that, for a valence orbital in an ion with large enough $Z$, $q \sim I_n$ as $\nu \rightarrow n$, the principal quantum number. The ratio $q_\textrm{norm} = -q/(Z^2\alpha^2 I_n)$ was computed for the $5s$ valence orbital along the Ag isoelectronic sequence; we reproduce the results (Fig.~1. of~\cite{berengut10prl}) as the filled circles in ~\Fig{fig:q5s_norm}, below.

Consider how \eref{eq:q_external_electron} might be modified for a hole. For our purposes, the hole state has the same quantum numbers as an electron in the closed shells. The hole state will have an enhanced sensitivity to $\alpha$-variation since the ionization potential is larger when there are more electrons in the outer shell, and is maximal for electrons in a closed shell. Indeed, $Z_a$ for an electron in closed shells would not be well described by $Z_a = Z_\textrm{ion}+1$ as it was previously; rather such an electron would spend approximately half its time closer to the nucleus than the other electrons in the same shell, and this leads to a larger relativistic shift. In such a case $q \sim I_n^{3/2}$ will better describe the system.

The open squares in \Fig{fig:q5s_norm} show the ratio $q_\textrm{norm}$ for the $5s$ orbital along the sequence $5s^2 4f^k$ with constant $Z_\textrm{ion}$, computed in the Dirac-Fock approximation.
The sequence starts at $5s^2$ ($k=0$) at $Z=60$, and ends at $5s^2 4f^{14}$ at $Z=74$. For all ions in the sequence $Z_\textrm{ion}=12$. Along this sequence the $4f$ orbital energies lie above the $5s$ orbital energies by less than a few percent -- in this sense the orbitals are in the same ``shell''.

\Fig{fig:q5s_norm} shows that there is indeed an extra contribution to the relativistic shift when the external orbital has many electrons in it: $q$ is increasing much faster than $I_n$. In the Hartree-Fock approximation for filled subshells, the ionization energy (to create a hole) is equal to the orbital energy. We see that a single hole in an otherwise filled external shell will have the highest sensitivity to variation of $\alpha$ possible given a particular $Z$ and $Z_a$. It is for this reason that we consider hole transitions as candidates for optical clocks.

To exploit the enhanced sensitivity to $\alpha$-variation afforded by hole transitions, we must find examples that have high $Z$, are highly ionized, and lie within the optical range. This is not trivial, since the difference in ionization potential between different shells increases as $\sim Z_a^2$, which rapidly takes any transition between the shells outside the range of lasers as $Z_a$ increases. To combat this tendency we must find particular examples near crossing points between the shells. The Coulomb crossing happens because in neutral atoms shells with larger angular momentum, $l$, can have energies significantly above shells with smaller momentum but larger principal quantum number, $n$ (for example, in neutral thorium the $5f$ orbital is above $7s$). On the other hand, in the hydrogenlike limit all shells with the same $n$ are nearly degenerate, regardless of angular momentum (e.g.~$E_{5f} = E_{5s}$). In this limit, a higher principal quantum number is necessarily associated with a larger energy. Therefore as $Z$ increases between the two limits there must be a crossing point where the two shells with different $n$ and $l$ have similar energies. Near such a point it may be possible to find hole-transitions within the range of lasers.

\section{Iridium}
\begin{figure}[tb]
\includegraphics[width=0.47\textwidth]{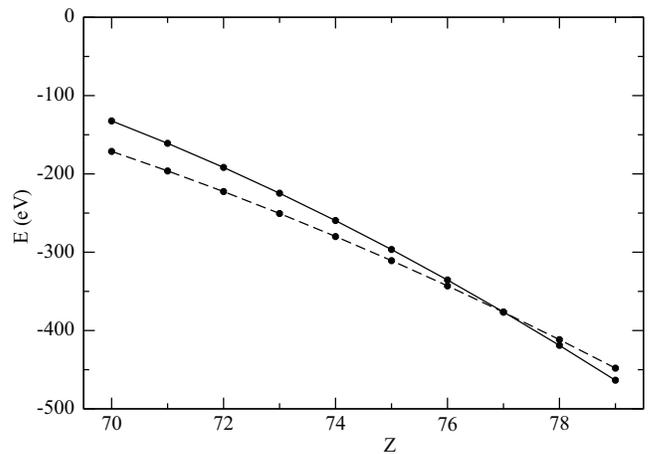}
\caption{\label{fig:E_N62} Dirac-Fock energies of $4f_{7/2}$ (solid) and $5s$ (dashed) orbitals in the Sm isoelectronic sequence, $N=62$, with closed external shells $5s^2 4f^{14}$.}
\end{figure}

In this \paper\ we identify two separate crossing points occurring in high-$Z$ ions. The first happens between the $4f^{14}$ and $5s^2$ shells in Sm-like ions (number of electrons $N=62$) with $Z\gtrsim 70$. In \Fig{fig:E_N62} we show the Dirac-Fock energies of these closed, outermost shells as a function of $Z$. The crossing is seen to occur around $Z=77$ (iridium). In the Dirac-Fock calculation a single hole has energy given by the ionization potential, $-E$. While the ionization potentials of both shells are around $380$~eV, the transition energy between configurations of Ir$^{16+}$ having the hole in the $4f$ level ($5s^2 4f^{13}$) and the hole in the $5s$ level ($5s\,4f^{14}$) will be very much less than this. Configuration interaction calculations are presented in \Tref{tab:Ir16+}. Note that for $Z$ below $70$ the $5p$ shell can begin to fill before the $4f$ shell. This is not necessarily a bad thing: for some $Z$ and some $N < 62$ there will be crossing between partly-filled $5p$ and $4f$ shells that may be exploited to find optical transitions, however we have not considered these many-electron cases in this work because no additional advantage is gained from more complex cases.

\begin{table}[tb]
\caption{Energy levels and sensitivity coefficients ($q$) for Ir$^{16+}$ (\cm).}
\label{tab:Ir16+}
\begin{ruledtabular}
\begin{tabular}{lcrr}
\multicolumn{1}{c}{Configuration} & $J$ &
\multicolumn{1}{c}{Energy} & 
\multicolumn{1}{c}{$q$} \\
\hline
$4f^{13}5s^2\ ^2\!F^o$ & 7/2 &        0 &        0 \\
                       & 5/2 &  25\,898 &  23\,652 \\
$4f^{14}5s\ ^2\!S$     & 1/2 &  37\,460 & 367\,315 \\
\end{tabular}
\end{ruledtabular}
\end{table}

\Tref{tab:Ir16+} and \Tref{tab:Ir17+} show configuration interaction (CI) calculations for low-lying levels in Ir$^{16+}$ and Ir$^{17+}$: the one-hole and two-hole cases, respectively. Details of the CI calculations have been presented elsewhere (see, e.g.~\cite{dzuba99pra,berengut04praB}). The $q$-values are calculated by repeating the entire calculation for $x=-0.01$ and $0.01$ and taking the gradient of the frequency, i.e.:
\[
q = \frac{\omega_{\delta x} - \omega_{-\delta x}}{2\,\delta x} \,,
\]
with $\delta x = 0.01$.
The resulting $q$ values are the largest ever seen in an atomic system.
Energy levels are accurate to 6000\,\cm, mostly due to two-electron excitations in the CI calculation not being fully saturated. However, the fine-structure splitting within terms is much more accurate than this, and the $q$-values are very stable and certainly correct to within 10\%.

\begin{table}
\caption{Energy levels and sensitivity coefficients ($q$) for Ir$^{17+}$ (\cm). Uncertainties between terms are at the level $\sim 6000\,\cm$, and in particular it is possible that the ground state is actually $4f^{14}\ ^1\!S_0$. However, since this level is metastable, it does not qualitatively affect our analysis of this system.}
\label{tab:Ir17+}
\begin{ruledtabular}
\begin{tabular}{lcrr}
\multicolumn{1}{c}{Configuration} & $J$ &
\multicolumn{1}{c}{Energy} & 
\multicolumn{1}{c}{$q$} \\
\hline
\multicolumn{4}{c}{Odd states} \\
$4f^{13}5s\ ^3\!F^o$   & 4 &        0 &       0 \\
$4f^{13}5s\ ^3\!F^o$   & 3 &   4\,838 &  2\,065 \\
$4f^{13}5s\ ^3\!F^o$   & 2 &  26\,272 & 24\,183 \\
$4f^{13}5s\ ^1\!F^o$   & 3 &  31\,492 & 25\,052 \\
\multicolumn{4}{c}{Even states} \\
$4f^{14}\ ^1\!S$     & 0 &   5\,055 &  367\,161 \\
$4f^{12}5s^2\ ^3\!H$ & 6 &  35\,285 & -385\,367 \\
$4f^{12}5s^2\ ^3\!F$ & 4 &  45\,214 & -387\,086 \\
$4f^{12}5s^2\ ^3\!H$ & 5 &  59\,727 & -362\,127 \\
$4f^{12}5s^2\ ^3\!F$ & 2 &  68\,538 & -378\,554 \\
$4f^{12}5s^2\ ^1\!G$ & 4 &  68\,885 & -360\,678 \\
$4f^{12}5s^2\ ^3\!F$ & 3 &  71\,917 & -362\,313 \\
$4f^{12}5s^2\ ^3\!H$ & 4 &  92\,224 & -339\,253 \\
$4f^{12}5s^2\ ^1\!D$ & 2 &  98\,067 & -363\,983 \\
$4f^{12}5s^2\ ^1\!J$ & 6 & 110\,065 & -364\,732 \\
$4f^{12}5s^2\ ^3\!P$ & 0 & 110\,717 & -372\,570 \\
$4f^{12}5s^2\ ^3\!P$ & 1 & 116\,372 & -362\,937 \\
\end{tabular}
\end{ruledtabular}
\end{table}

Let us take two particular examples from Ir$^{17+}$, the two-hole case. Trapping and cooling of this ion may be facilitated by any of the E1 transitions from the ground state $4f^{13}5s\ ^3\!F_4^o$, such as $4f^{12}5s^2\ ^3\!F_4$ at $\omega \sim 45\,000\,\cm$. While this transition itself has an extremely large $q$-value, and could therefore in principle be used to test $\alpha$-variation, E1 transitions tend to be too broad for precise measurements. On the other hand the transition to $4f^{12}5s^2\ ^3\!H_6$ with frequency $\sim 35\,000\,\cm$ could be induced via hyperfine mixing with its fine-structure partner $^3H_5$, and this transition would have a strongly reduced linewidth (it may also go as E3/M2). The most abundant isotope of iridium is $^{193}$Ir, which is stable and has nuclear spin $I = 3/2$. Therefore the hyperfine-induced transitions are always available.

The $4f^{14}\ ^1\!S_0$ line is interesting because the $\alpha$-sensitivity acts in the opposite direction ($q$ is large and positive). However, when trying to use this line one expects to encounter the problem of too little line strength. The best option here is to use the transition from the very metastable $4f^{14}\ ^1\!S_0$ to $4f^{13}5s\ ^3\!F_2^o$ ($\omega \approx 21\,000\,\cm$). This is an M2/E3 transition, and at first seems too slow for existing lasers. However, due to the hyperfine mixing with $J=1$ states such as the $116\,000\,\cm$ $4f^{12}5s^2\ ^3\!P_1$ level, this transition may also proceed via hyperfine-induced E1 transition. Odd-parity $J=1$ states also exist; hyperfine mixing with these states would be dominated by the configuration $4f^{12}5s5p$ at energies $\sim 500\,000\,\cm \sim 60$\,eV, which is still far below the continuum of such a highly-ionized system.

The best current limit on terrestrial time-variation of $\alpha$ comes from an experiment comparing Hg$^+$ and Al$^+$ clocks~\cite{rosenband08sci}, with $\Delta q = 57\,000\,\cm$~\cite{dzuba99pra}. By contrast, an experiment that compared the two transitions we have examined ($4f^{14}\ ^1\!S_0$ to $4f^{13}5s\ ^3\!F_2^o$ and $4f^{13}5s\ ^3\!F_4^o$ to $4f^{12}5s^2\ ^3\!H_6$) would have a total sensitivity to $\alpha$-variation of $\Delta q \approx 730\,000\,\cm$. The largest $q$ ever found in a single atomic transition occurs in $4f^{14}\ ^1\!S_0$ to $4f^{12}5s^2\ ^3\!F_2$, however in this case the upper level has an E1-decay mechanism and may be too broad.

\section{Tungsten}
\begin{figure}[tb]
\includegraphics[width=0.47\textwidth]{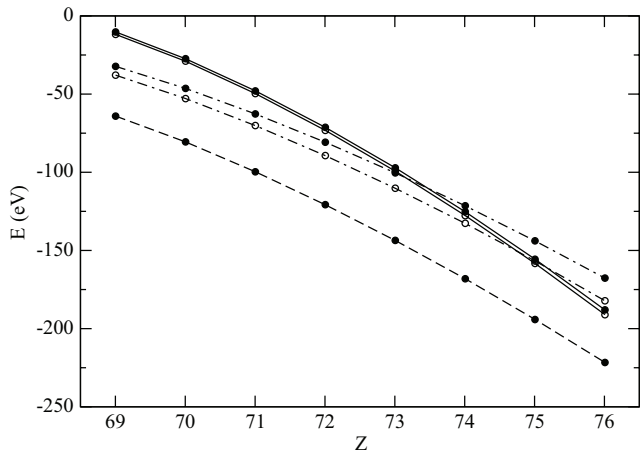}
\caption{\label{fig:E_N68} Dirac-Fock energies of $4f$ (solid), $5s$ (dashed), and $5p$ (dot-dashed) orbitals in the Er isoelectronic sequence, $N=68$, with closed external shells $5s^2 5p^6 4f^{14}$. The orbital fine-structure splitting is indicated, with closed circles for $4f_{7/2}$ and $5p_{3/2}$ and open circles for the $4f_{5/2}$ and $5p_{1/2}$ orbitals.}
\end{figure}

The second crossing point we consider happens when the filled $4f^{14}$ and $5p^6$ shells cross in Er-like ions ($N=68$). Because of the relatively large splitting between the $5p_{3/2}$ and $5p_{1/2}$ orbitals (compared to the much smaller fine-structure splitting of the $4f$ orbitals), this crossing point is more spread out, somewhere between $Z=73$ and~75 (see \Fig{fig:E_N68}). We focus on $Z=74$ because tungsten is a well-studied element due to its use in plasma diagnostics (see, e.g.~\cite{clementson10jpb}), and hence may be of higher interest to experimenters. CI calculations for the one-hole and two-hole systems, W$^{7+}$ and W$^{8+}$ respectively, are presented in Tables~\ref{tab:W7+} and~\ref{tab:W8+}.

\begin{table}[tb]
\caption{Energy levels and sensitivity coefficients ($q$) relative to the ground state for W$^{7+}$ (\cm).}
\label{tab:W7+}
\begin{ruledtabular}
\begin{tabular}{lcrbr}
\multicolumn{1}{c}{Configuration} & $J$ &
\multicolumn{2}{c}{Energy} & 
\multicolumn{1}{c}{$q$} \\
 & & \multicolumn{1}{c}{This work} & \multicolumn{1}{c}{\cite{kramida09adndt}} & \\
\hline
$4f^{13}5p^6\ ^2\!F^o$ & 7/2 &       0 &       0 &       0 \\
                       & 5/2 & 18\,199 & 17\,440 & 16\,462 \\
$4f^{14}5p^5\ ^2\!P^o$ & 3/2 &  4\,351 &     800 (700) &  87\,544 \\
                       & 1/2 & 93\,908 & 87\,900 (700) & 200\,269 \\
\end{tabular}
\end{ruledtabular}
\end{table}

\begin{table}[tb]
\caption{Energy levels and sensitivity coefficients ($q$) for W$^{8+}$ (\cm). Energies between terms are uncertain at the level $\sim 6000\,\cm$.}
\label{tab:W8+}
\begin{ruledtabular}
\begin{tabular}{lcrr}
\multicolumn{1}{c}{Configuration} & $J$ &
\multicolumn{1}{c}{Energy} & 
\multicolumn{1}{c}{$q$} \\
\hline
$4f^{14}5p^4\ ^3P$ & 2 &       0 &        0 \\
$4f^{13}5p^5\ ^3F$ & 4 &  6\,075 &  -81\,564 \\
$4f^{13}5p^5\ ^3G$ & 3 &  6\,357 &  -81\,480 \\
$4f^{13}5p^5\ ^3G$ & 5 & 11\,122 &  -82\,880 \\
$4f^{13}5p^5\ ^3F$ & 3 & 21\,905 &  -66\,489 \\
$4f^{13}5p^5\ ^3D$ & 2 & 23\,276 &  -66\,985 \\
$4f^{13}5p^5\ ^3F$ & 2 & 28\,112 &  -66\,124 \\
$4f^{14}5p^4\ ^1S$ & 0 & 29\,810 &   12\,735 \\
$4f^{13}5p^5\ ^3G$ & 4 & 34\,884 &  -65\,896 \\
$4f^{13}5p^5\ ^3D$ & 1 & 36\,497 &  -65\,946 \\
$4f^{12}5p^6\ ^3H$ & 6 & 56\,416 & -179\,990 \\
$4f^{12}5p^6\ ^3F$ & 4 & 65\,008 & -181\,419 \\
$4f^{12}5p^6\ ^3H$ & 5 & 73\,188 & -164\,139 \\
$4f^{12}5p^6\ ^1G$ & 4 & 80\,551 & -162\,573 \\
$4f^{12}5p^6\ ^3F$ & 2 & 82\,424 & -177\,137 \\
$4f^{12}5p^6\ ^3F$ & 3 & 83\,315 & -164\,072 \\
\end{tabular}
\end{ruledtabular}
\end{table}

Previous studies extrapolating from the measured W$^{6+}$ spectrum~\cite{sugar75pra} have noted that the $4f$ and $5p$ orbitals compete for the ground state in W$^{7+}$~\cite{kramida09adndt}. We present these calculations in \Tref{tab:W7+} but repeat the warnings of~\cite{kramida09adndt} that the accuracy may be low. The fine-structure splittings are more accurate.

Both W$^{7+}$ and W$^{8+}$ have several M1 and E2 transitions that may be suitably narrow to be of use in studies of $\alpha$-variation. For example, the E2 transition from the ground state of W$^{8+}$, $4f^{14}5p^4\ ^3P_2$, to $4f^{12}5p^6\ ^3F_4$ has energy $\omega \sim 65\,000\,\cm$ and $q = -181\,000\,\cm$. This transition may be compared to the $4f^{14}5p^4\ ^1S_0$ transition, or to any line with a positive or small $q$ value in another ion, to give a sensitivity to $\alpha$-variation which would still be the largest ever utilised. By comparing instead to the $4f^{13}5p^6\ ^2\!F^o_{5/2}$ to $4f^{14}5p^5\ ^2\!P^o_{1/2}$ E2 transition in W$^{7+}$, one could achieve $\Delta q \approx 365\,000\,\cm$.

\section{Conclusion}
We have shown that level-crossings between filled shells provide an opportunity to find hole-transitions within laser range in highly-charged ions. These hole transitions can have hugely enhanced sensitivity to $\alpha$-variation when compared with transitions used today. As atomic spectroscopy in electron-beam ion traps continues to improve (see, e.g.~\cite{draganic03prl,crespo08cjp,hobein11prl} and review~\cite{beiersdorfer09pscr}) we expect that hole-transitions in multiply-charged ions may become good systems for searching for time-variation of $\alpha$. We have found candidates in two different systems that may be of interest, and suggest that others may exist along other isoelectronic sequences. Level crossings also allow us to identify optical-range E1 transitions such as those we have found in Ir$^{17+}$ that may be of importance in laser-trapping and cooling of these ions.

\acknowledgments

This work was supported by the Australian Research Council. Supercomputer time was provided by an award under the Merit Allocation Scheme on 
the NCI National Facility at the Australian National University.

\bibliography{references}

\end{document}